\documentclass[fleqn,twoside]{article}
\usepackage{espcrc2,psfrag,epsfig,times}

\usepackage{graphicx}
\usepackage[figuresright]{rotating}

\newcommand{\AmS}{{\protect\the\textfont2
  A\kern-.1667em\lower.5ex\hbox{M}\kern-.125emS}}

\newcommand{\cha}{\tilde{\chi}}
\newcommand{\neu}{\tilde{\chi}^0}

\title{Reconstruction of Fundamental SUSY Parameters%
        \thanks{Expanded version of contributions to the proceedings
                of ICHEP.2002, Amsterdam (Nederland), 
                and LCWS.2002, Jeju Island (Korea),
                by the SUSY Collaboration of the ECFA/DESY LC Workshop
                .} \hfill
        $^{^{^{^{^{^{\mbox{\small DESY--02--175}}_{
                           \mbox{\small ZU-TH 21/02}}}}}}}$}
        \author{P.M. \hspace{-1pt}Zerwas\address[DESY]{Deutsches 
                                Elektronen--Synchrotron DESY,
                                D-22603 Hamburg, Germany},
        J. \hspace{-1pt}Kalinowski\address{Insitute of Theoretical Physics, Warsaw
                        University, 00681 Warsaw, Poland},
        A. \hspace{-1pt}Freitas\address{Fermi National Accelerator Laboratory, Batavia, IL
                        60510-500, USA},
        G.A. \hspace{-1pt}Blair\addressmark[DESY]$^{,}$\address{Royal Holloway and Bedford
                        New College, University of London, UK},
        S.Y. \hspace{-1pt}Choi\address{Chonbuk National University, Chonju 561-756, Korea},
        H.U. \hspace{-1pt}Martyn\address{I.~Physik.~Institut, 
                                        RWTH Aachen, D-52056 Aachen, Germany},
        G. \hspace{-1pt}Moortgat-Pick\addressmark[DESY],
        W. \hspace{-1pt}Porod\address{Inst.~Theor.~Physik, 
                         Universit\"at Z\"urich, CH-8057 Z\"urich, Switzerland}
}

\begin{document}

\begin{abstract}
We summarize methods and expected accuracies in determining the basic
low-energy SUSY parameters from experiments at future e$^+$e$^-$ linear
colliders in the TeV energy range, combined with results from LHC. 
In a second step we demonstrate how, based on this set of parameters,
the fundamental supersymmetric theory can be reconstructed at high scales
near the grand unification or Planck scale. These analyses have been
carried out for minimal supergravity 
[confronted with GMSB
for comparison], and for a string effective theory.
\end{abstract}

\maketitle

\section{Introduction}
\noindent
Standard particle physics is characterized by energy scales of order 100
GeV. However the roots for all the phenomena we observe experimentally in
this range, may go as deep as the Planck length of $10^{-33}$~cm, 
equivalently
to energies near the Planck scale
$\Lambda_{\rm PL} \sim 10^{19}$~GeV or the grand unification [GUT] scale
$\Lambda_{GUT} \sim 10^{16}$~GeV.
Supersymmetry [SUSY] \cite{WsZu,NilHab} provides us with a
stable bridge \cite{witten:81} between these two vastly different energy 
regions.
We expect the origin of supersymmetry breaking at the high scale. The
breaking mechanism may have its base in a hidden world
connected by gravity with our own eigen-world in which we observe the
SUSY phenomena. This scenario is realized in minimal supergravity
[mSUGRA], cf.~Ref.\cite{sugra}. 
Supersymmetry breaking may microscopically
be generated in string theories, leading to effective field theories,
Ref.\cite{Binetruy:2001md}, in which the breaking mechanism is encoded 
in local fields in
four dimensions and transferred by their interactions with matter and gauge
fields to the observed phenomena in our eigen-world.

To study the fundamental structure of theories at scales as high as the
Planck scale, only a few tools are available to us. We may use proton
decay and related phenomena, likely the neutrino sector and
quark/lepton mass-matrix textures, as well as the cosmology of the early
universe. The total of information, however, remains scarce and the
methods are sometimes rather indirect. On the other hand, a
rich corpus
of information on physics near the Planck scale may become available
from the well-controlled extrapolation of fundamental parameters 
measured with high precision at laboratory energies.
Such extrapolations extend over 13 to 16
orders of magnitude. Despite this huge distance, they can
be carried out in a stable way in supersymmetric theories. To this purpose
renormalization group techniques are exploited, by which parameters
are transported from low to high scales based on nothing but measured
quantities in laboratory experiments. This procedure has very
successfully been pursued for the three electroweak and strong gauge couplings.
Universality of these three couplings is the solid base of the grand
unification hypothesis. Small deviations from nearly perfect regularities
can be explored to investigate genuine high-scale structures. In this way
a telescope can be built to physics near the Planck scale.

The method can be expanded to a large ensemble of supersymmetry
parameters~\cite{Blair:2000gy,KaneG} -- the soft SUSY breaking parameters: 
gaugino and scalar
masses, as well as trilinear couplings. We have analyzed this procedure for
two examples. The first, minimal supergravity, is characterized 
by a naturally high degree of
regularity near the grand unification scale. [The pattern of the
extrapolated mSUGRA parameters is subsequently confronted 
with gauge mediated supersymmetry breaking GMSB \cite{gmsb}
to demonstrate sensitivity and uniqueness]. 
In a second step, the parameters of effective field 
theories based on orbifold compactification of the heterotic string, 
are analyzed. This bottom-up approach, formulated by means of
the renormalization group, makes use of the low-energy measurements to
the maximum extent possible and it reveals the quality with which the
fundamental theory at the high scale can be reconstructed in a transparent
way.

The basic structure in this approach is assumed to be essentially 
of desert type.
Nevertheless, the existence of intermediate scales is not precluded.
An important example is provided by the left-right extension of mSUGRA
incorporating the seesaw mechanism for the masses of right-handed neutrinos
at scales beyond $10^{10}$~GeV. 

High-quality experimental data are necessary in this context, that should
become available by future lepton colliders~\cite{LC,BWZ} in a unique way.
We shall 
study how well such a program can be realized at e$^+$e$^-$ linear colliders,
ranging from LC in the 1~TeV range [such as TESLA] 
to multi-TeV energies [such as CLIC], 
and combined with information that will be extracted from LHC analyses
[see also Ref.\cite{NO}].

After discussing first the measurements of the basic SUSY parameters at
the low scale, we will summarize in the second step the results expected
from the reconstruction of the fundamental supersymmetric theory at the
grand unification or Planck scale in the two scenarios defined above.

\section{Minimal Supergravity}
\noindent
Supersymmetry is broken in mSUGRA in a hidden sector and the breaking is
transmitted to our eigen-world by gravity~\cite{sugra}. 
This mechanism suggests, yet does not enforce,
the universality of the soft SUSY breaking parameters
at a scale which we will identify with the unification scale 
for the sake of simplicity.

\begin{figure}[p]
\begin{center}
\setlength{\unitlength}{1mm}
\begin{picture}(75,70)
\psfrag{h}{\Huge $h^0$}
\psfrag{J}{\Huge $H^0$}
\psfrag{Z}{\Huge $\,A^0$}
\psfrag{I}{\Huge $H^\pm$}
\psfrag{N}{\Huge $\neu_1$}
\psfrag{M}{\Huge $\neu_2$}
\psfrag{O}{\Huge $\cha^\pm_1$}
\psfrag{Y}{\Huge $\neu_3$}
\psfrag{X}{\Huge $\tilde{\tau}_1$}
\psfrag{T}{\Huge $\tilde{\tau}_2$}
\psfrag{R}{\Huge $\tilde{e}_{\rm R}$}
\psfrag{L}{\Huge $\tilde{e}_{\rm L}$}
\psfrag{V}{\Huge $\tilde{\nu}_\tau$}
\psfrag{U}{\Huge $\tilde{\nu}_e$}
\psfrag{B}{\Huge $\tilde{t}_1$}
\psfrag{C}[r][r]{\Huge $\tilde{u}_{\rm R}, \tilde{d}_{\rm R}$}
\psfrag{D}{\Huge $\tilde{u}_{\rm L}$}
\psfrag{F}[r][r]{\Huge $\tilde{t}_2$}
\psfrag{Q}[r][r]{\Huge $\neu_4, \cha^\pm_2\!\!\!\!\!\!$}
\put(0,0){%
\epsfig{figure=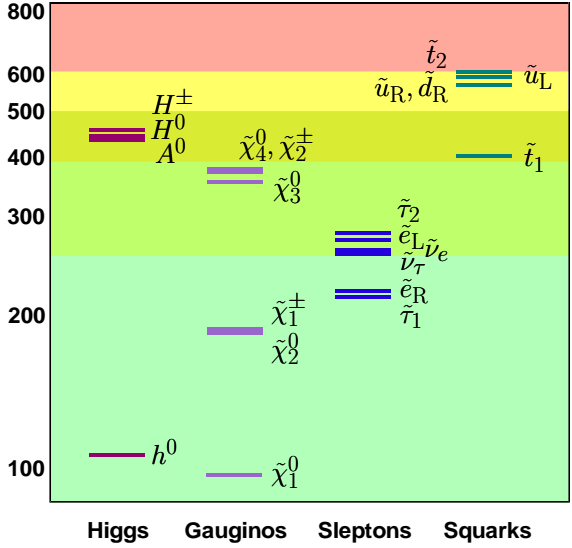, angle=0, width=7.5cm, viewport=32 0 540 505, clip=true}
}
\put(6,74.5){\line(68,0){68.5}}
\end{picture}
\end{center}
\vspace{-12mm}
\caption{\it The spectrum of Higgs particles,
             gauginos and scalar particles in the mSUGRA scenario
             {\it mod.}SPS\#1A; masses in {\normalfont GeV}.}
\label{fig:freitas}
\end{figure}

\begin{figure}[tb]
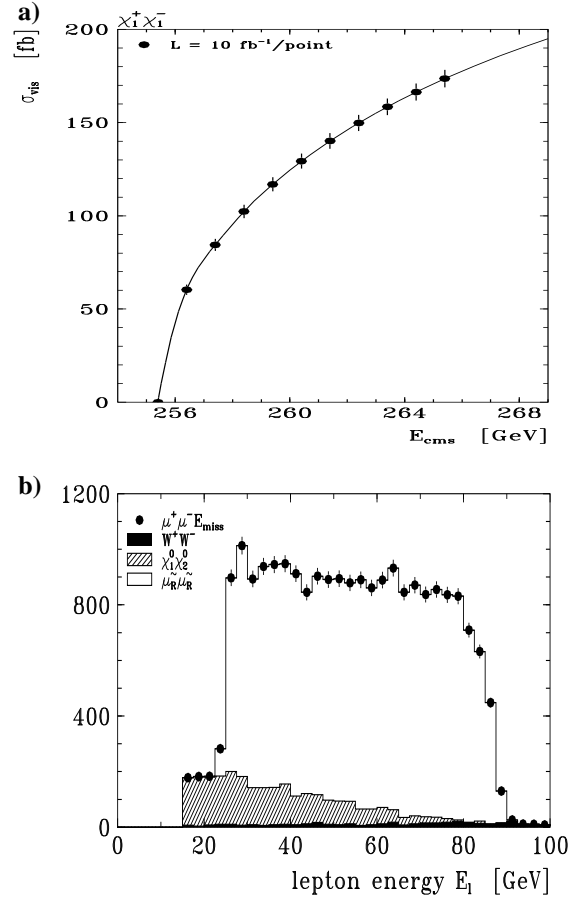

\setlength{\unitlength}{1mm}
\begin{picture}(70,120)
\put(-3,60){
\mbox{\includegraphics[height=9.cm,width=7cm,angle=90,viewport=0 340 555 781]{c11scan.eps}}}
\put(-3,0){
\mbox{\includegraphics[height=8.cm,width=7cm,angle=90]
{mur132.emu.w320.eps}}}
\put(0,120){\mbox{\bf a)}}
\put(0,57){\mbox{\bf b)}}
\end{picture}
\vspace{-10mm}
\caption{\it a) Threshold excitation in chargino pair production, b) $\mu$
energy spectrum in $\tilde{\mu}_{\rm R}$ decays; see Ref.\cite{blair2}.}
\label{fig:martyn}
\end{figure}

The typical form of the mass spectrum in mSUGRA scenarios can be
exemplified by the Snowmass point SPS\#1A~\cite{Allanach:2002nj}, 
slightly modified for
illustrative purpose by increasing the scalar mass parameter,
as shown in Fig.~\ref{fig:freitas}. 
This reference point is
compatible with all known constraints from precision data and search
experiments. Moreover, it does not require an excessive amount of
fine-tuning either for the electroweak parameters or for cold dark
matter. In this scenario, the non-colored gauginos and scalar leptons can
be produced at LC while squarks and gluino parameters can be measured
from LHC and CLIC. High-precision analyses of the light and heavy Higgs
sectors need the operation of LC and CLIC.
Masses can best be obtained in threshold scans at e$^+$e$^-$ 
colliders~\cite{blair2}. The
excitation curves for chargino production in S-waves \cite{Choi:2000ta} 
rise steeply with the velocity
of the particles near the threshold and thus are very sensitive to
their masses; the same is true for mixed-chiral selectron pairs in $e^+e^-$ 
and for diagonal pairs in $e^-e^-$
collisions. Other scalar fermions as well as neutralinos are produced
generally in P-waves, with a somewhat less steep threshold behavior
proportional to the third power of the velocity~\cite{freitas}. 
Additional information,
in particular on the lightest neutralino $\tilde{\chi}^0$, can be extracted from
decay spectra. Two characteristic examples are depicted in 
Fig.~\ref{fig:martyn}(a) and (b).
A selection of parameters 
combined from TESLA, LHC and CLIC
measurements is collected in Tab.~\ref{tab:masses}.%

\renewcommand{\arraystretch}{1.1}
\begin{table}[p]
\begin{center}
\begin{tabular}{|c||r||l|}
\hline
 & Meas.+ Errors & \\ \hline\hline
$\tilde{\chi}^\pm_1$ &$183.05\pm0.15$   &LC\\
$\tilde{\chi}^\pm_2$ & $383.28\pm 0.28$   &LC\\
$\tilde{\chi}^0_1$ & $97.86\pm 0.20$   &LC\\
$\tilde{\chi}^0_2$ & $184.65\pm 0.30$   &LC\\ \hline
$\tilde{e}_R$ & $224.82\pm 0.15$   &LC\\
$\tilde{e}_L$ & $269.09\pm 0.28$   &LC\\
$\tilde{u}_R$ & $572\pm 10\phantom{.0}$    &LHC+CLIC\\
$\tilde{u}_L$ & $589\pm 10\phantom{.0}$   &LHC+CLIC  \\ \hline
$\tilde{g}$ &  $598\pm 10\phantom{.0}$   &LHC  \\
\hline
$h^0$ & $113.38\pm0.05$ & LHC+LC\\
$A^0$ & $435.5\pm1.5\phantom{0}$ & CLIC \\
\hline 
\end{tabular}\\
\end{center}
\caption{\it A sample of masses and expected accuracies for {{\it mod.}SPS\#1A};
             masses in {\normalfont GeV}. The mSUGRA point probed 
is defined by the parameters $M_0 = 200$~{\normalfont GeV}, 
$M_{1/2} = 250$~{\normalfont GeV}, $A_0$ = $-100$~{\normalfont GeV}, 
$\tan \beta = 10$, 
and $\mathrm{sign}(\mu) = (+)$.\newline}
\label{tab:masses}
\end{table}

\begin{figure}[tb]
\begin{center}
\includegraphics[width=15pc]{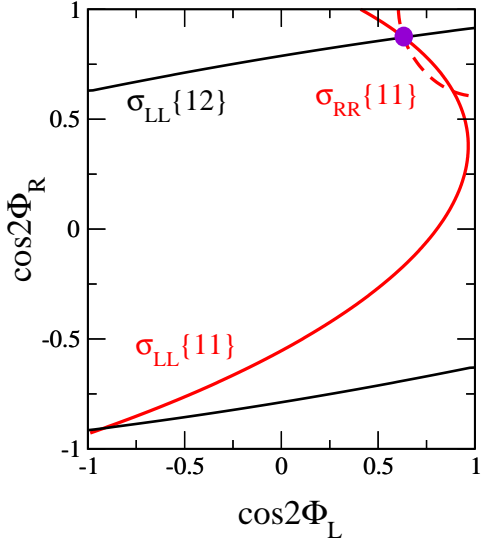}
\end{center}
\vspace{-12mm}
\caption{\it Determining the chargino mixing angles in measurements
             of chargino production cross sections \{ij\} for polarized
             beams.}
\label{fig:choi}
\end{figure}

Mixing parameters must be obtained from measurements of cross
sections, in particular from the production of chargino pairs and
neutralino pairs~\cite{Choi:2000ta}, both in diagonal or mixed form: 
$e^+e^- \rightarrow {\tilde{\chi}^+_i}{\tilde{\chi}^-_j}$
[$i$,$j$ = 1,2] and ${\tilde{\chi}^0_i} {\tilde{\chi}^0_j}$ [$i$,$j$ = 1,$\dots$,4]. 
The production cross sections for
charginos are binomials of $\cos\,2\phi_{L,R}$, the mixing angles
rotating current to mass eigenstates. Using polarized electron
and positron beams, the cosines can be determined in a model-independent
way, Fig.~\ref{fig:choi}.
Similarly, measurements of the cross sections for sfermion production
with polarized beams are needed to determine mixing angles and
trilinear couplings in this sector.


Based on this high-precision information, the fundamental SUSY
parameters can be extracted at low energy in analytic form. To lowest
order:
\begin{eqnarray}
\left|\mu\right|&=&M_W[\Sigma + \Delta[\cos2\phi_R+\cos2\phi_L]]^{1/2}
\nonumber\\
\mbox{sign}(\mu)&= &[ \Delta^2
                   -(M^2_2-\mu^2)^2-4m^2_W(M^2_2+\mu^2) \nonumber \\
 & &                   -4m^4_W\cos^2 2\beta]/8 m_W^2M_2|\mu|\sin2\beta 
\nonumber\\
M_2&=&M_W[\Sigma - \Delta(\cos2\phi_R+\cos2\phi_L)]^{1/2}\nonumber\\
|M_1|&=& \left[ \textstyle \sum_i m^2_{\tilde{\chi}_i^0}  
                 -M^2_2-\mu^2-2M^2_Z\right]^{1/2}
\nonumber
\end{eqnarray}\pagebreak[2]
\begin{eqnarray}
|M_3|&=&m_{\tilde{g}} \nonumber\\
\tan\beta&=&\left[\frac{1+\Delta (\cos 2\phi_R-\cos 2\phi_L)}
           {1-\Delta (\cos 2\phi_R-\cos 2\phi_L)}\right]^{1/2} 
\label{eqn:basicLE}
\end{eqnarray}
where $\Delta = (m^2_{\tilde{\chi}^\pm_2}-m^2_{\tilde{\chi}^\pm_1})/(4M^2_W)$
and 
$\Sigma =  (m^2_{\tilde{\chi}^\pm_2}+m^2_{\tilde{\chi}^\pm_1})/(2M^2_W) -1$.
The signs of $M_{1,3}$ with respect to $M_2$ 
will follow from measurements of
the cross sections for ${\tilde{\chi}} {\tilde{\chi}}$ production and gluino processes. 
In practice 
one-loop corrections to the mass relations have been used to improve on the accuracy.

Accuracies expected for the parameters in the reference point
{\it mod.}SPS\#1A are shown
in Tab.~\ref{tab:parvalues_a}. While the LC based errors are typically 
at the per-mille level, the others turn out to be in the per-cent range.

\begin{table}[tb]
\begin{center}
\setlength{\tabcolsep}{2pt}
\begin{tabular}{|c||c|c|}
\hline
 &  Exp.~Input &  GUT Value \\ \hline   \hline
 $M_1$ & 102.31 $\pm$  0.25 &  $250.00 \pm  0.33$ \\
 $M_2$ &  192.24 $\pm$  0.48      &  $250.00 \pm  0.52$ \\
 $M_3$ & \phantom{0}586  $\pm$  12   &  $250.0    \pm   5.3$  \\ \hline
$\mu$         & 358.23  $\pm$ 0.28     &  $355.6 \pm  1.2    $  \\ \hline
 $M^2_{L_1} $ & $( 6.768  \pm  0.005)\cdot 10^4$
                &  $(3.99  \pm  0.41) \cdot 10^4$  \\
 $M^2_{E_1} $ & $(4.835  \pm  0.007) \cdot 10^4$
  &  $(4.02  \pm  0.82)  \cdot 10^4 $ \\
 $M^2_{Q_1} $ &  $(3.27 \pm  0.08)\cdot 10^5$
               &  $(3.9  \pm  1.5) \cdot 10^4$ \\
  $M^2_{U_1} $ &  $(3.05 \pm  0.11)\cdot 10^5$
               &  $(3.9  \pm  1.9) \cdot 10^4$ \\\hline
 $M^2_{H_1} $ &  $\phantom{-}(6.21 \pm  0.08)\cdot 10^4$  &
  $ (4.01  \pm  0.54)  \cdot 10^4 $ \\
 $M^2_{H_2}$~ &  $(-1.298 \pm 0.004)\cdot 10^5$ &
  $(4.1  \pm  3.2) \cdot 10^4 $\\
 $A_t $ & $-446 \pm 14\phantom{0\cdot 10^4}$   &  $-100 \pm 54\phantom{-0\cdot
 10^4}$   \\ \hline
 $\tan\beta$ & $\phantom{-}9.9 \pm  0.9\phantom{\rule{0mm}{0mm}\cdot 10^4}$ & --- \\ \hline
\end{tabular}
\end{center}
\label{tab:parvalues_a}
\caption{\it Reconstructed gaugino and scalar mass parameters
             for {{\it mod.}SPS\#1A} at the electroweak scale and RG
             extrapolation to the GUT scale; all 
             masses in {\normalfont GeV}.}
\end{table}

It should be noted that knowledge of the low chargino/neutralino spectrum
${\tilde{\chi}^{\pm}_1}$ and $\tilde{\chi}^0_{1,2}$ is sufficient to carry out such 
an analysis. On the other
hand, Higgs couplings or polarization effects must be used in addition
to determine the Higgs parameter $\tan\beta$ for large values with sufficient
accuracy.

The evolution to the high scale is governed by solutions of the
renormalization group equations~\cite{RGE2}:\pagebreak
\begin{center}
\begin{tabular}{lcl}
 gauge couplings &:&  $\alpha_i = Z_i \, \alpha_U$  \\[.2ex]
  gaugino masses &:& $M_i = Z_i \, M_{1/2}$  \\[.2ex]
 scalar masses &:&   $M^2_{\tilde j} = M^2_0 + c_j M^2_{1/2}$ \\[.2ex]
& &       $\phantom{M^2_{\tilde j} = }
 + \sum_{\beta=1}^2 c'_{j \beta} \Delta M^2_\beta$   \\[.2ex]
  trilinear  couplings &:&  $A_k = d_k A_0   + d'_k M_{1/2}$  
\end{tabular}
\end{center}
\pagebreak[2]
The $Z$ transporter is given by 
$Z_i^{-1} =1 + b_i \, \alpha_U / (4 \pi)$ $\times \log(M_U / M_Z)^2$
with $b[SU_3, SU_2, U_1] = -3$, 1, $33 / 5$.  The $c$, $d$ 
coefficients as well as
the shifts $\Delta M^2_\beta$  depend on the high-energy 
parameters to be calculated,
so that complicated implicit equations emerge -- partly with
solutions of low sensitivity. In practice the transport
equations have been solved to two loops.

\begin{figure}[!t]
\setlength{\unitlength}{1mm}
{\bf a)}~~~ $1/M_i$~[GeV$^{-1}$]\\
\put(55,-3){$Q$ [GeV]}
\epsfig{figure=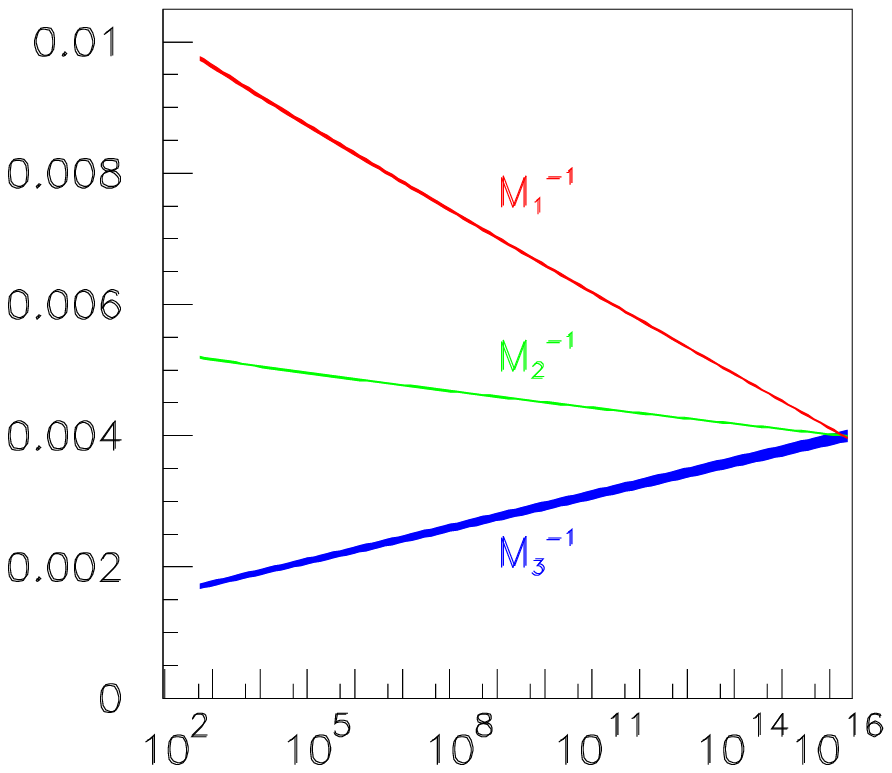,width=7cm,viewport=10 290 270 520}\\
{\bf b)}~~~ $M_j^2$ [GeV$^2$]\\
\put(55,-3){$Q$ [GeV]}
\epsfig{figure=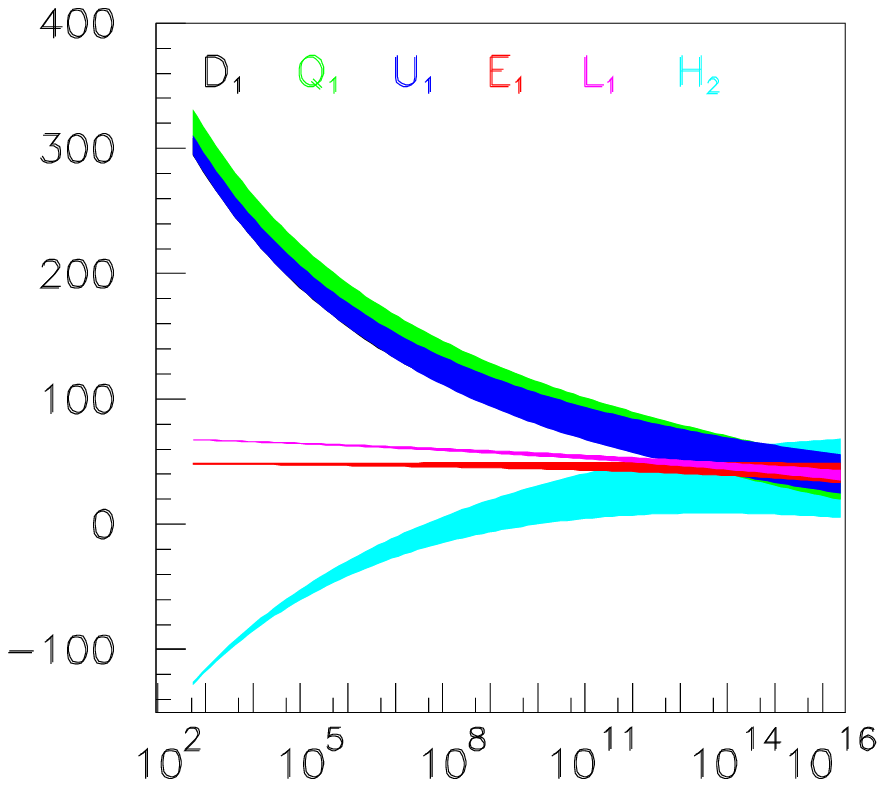,width=7cm,viewport=10 290 270 520}%
\vspace{-6mm}
\caption{{\bf mSUGRA:} {\it  Evolution, from low to high scales, of 
a) gaugino mass parameters, b) first-generation sfermion mass parameters squared
and the Higgs mass parameter $M^2_{H_2}$.   The mSUGRA point probed 
is defined in Tab.~\ref{tab:masses}.
[The widths of the bands indicate the 1$\sigma$ CL.]
}}
\label{fig:sugra}
\end{figure}

Examples for the gaugino masses and the scalar masses of the first two
generations are depicted in Fig.~\ref{fig:sugra}. 
Moving down from the universality point,
the mass squared 
of the Higgs field $H_2$ crosses to
negative values at a scale
of order $10^4$ TeV, in accordance with radiative symmetry breaking in mSUGRA.

\begin{table}[tb]
\begin{center}
\begin{tabular}{|c||c|c|}
\hline
                &  Ideal           & Exp. Error \\ \hline  \hline
$M_U$           & $2\cdot 10^{16}$ &  $1.6  \cdot 10^{14}$       \\
$\alpha_U^{-1}$ &   24.361          &     0.007     \\ \hline
$M_\frac{1}{2}$ & 250              & 0.08     \\
$M_0$           & 200              & 0.09     \\
$A_0$           & -100             & 1.8      \\  \hline
$\mu$           & 358.23           & 0.21     \\
$\tan\beta $    &  10              & 0.1      \\  \hline
\end{tabular}
\end{center}
\caption[]{ \it Comparison of the ideal parameters with the
experimental expectations for the particular mSUGRA reference
point analyzed in this report; all mass parameters are given 
in units of $GeV$.}
\label{tab:parvalues_b}
\end{table}%
In an overall-fit, based on the universality hypothesis
{\it per se}, we observe 
accuracies of order per-mille in the gaugino sector while being
order per-cent in the scalar sector, cf. Tab.~\ref{tab:parvalues_b}.
\pagebreak[4]

\begin{figure}[tb]
\setlength{\unitlength}{1mm}
\begin{center}
\begin{picture}(60,55)
\put(-1,-7){
\mbox{\includegraphics[height=6cm,width=6cm,angle=0]{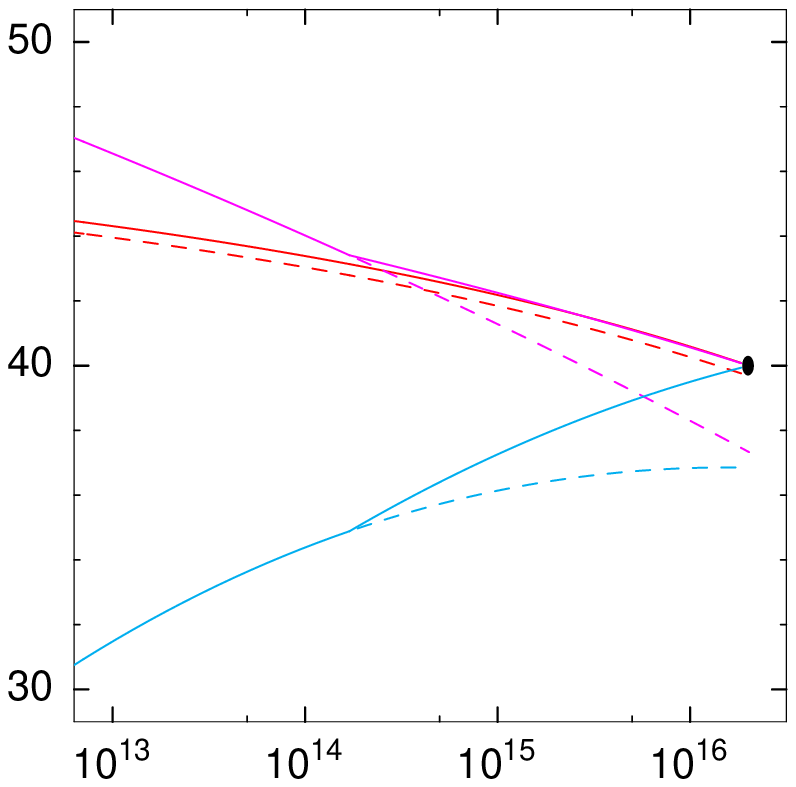}}}
\put(0,55){\mbox{$M^2_{\tilde\jmath}$~[$10^3$ GeV$^2$]}}
\put(50,-10){\mbox{$Q$~[GeV]}}
\put(8,43){\mbox{\small $M^2_{L_3}$}}
\put(8,31){\mbox{\small $M^2_{E_3}$}}
\put(8,10){\mbox{\small $M^2_{H_2}$}}
\put(27.8,6){\mbox{\small $\nu_{R_3}$}}
\put(27,6){\vector(0,1){5}}
\put(25,45){\mbox{\small ----- $\,\nu_R$, $\tilde{\nu}_R$ included}}
\put(25,41){\mbox{\small - - - $\nu_R$, $\tilde{\nu}_R$ excluded}}
\end{picture}
\end{center}
\caption{{\bf LR-SUGRA with $\nu_R$:} {\it  
Evolution of third-generation slepton mass parameters and the Higgs mass parameter
$M_{H_2}^2$. The mSUGRA point probed is characterized by the same parameters
as before while the $\nu_{R_3}$ scale is taken close to $10^{14}$ {\normalfont
GeV}.}}
\label{fig:lrsugra}
\end{figure}

\vspace{3mm}
\noindent
{\underline{\it Left-Right SUGRA:}} The universal SUGRA model can readily be
extended to a left-right symmetric theory [as suggested by non-zero neutrino
masses] if the SO(10) unification scale is located in between the SU(5) and 
the Planck scale. The first and second generation are not changed. Owing to the 
enhanced Yukawa coupling, the seesaw scale above $10^{10}$~GeV however is felt
in the evolution of the third generation -- albeit with weak sensitivity.
This effect is evident from Fig.~\ref{fig:lrsugra} in which the evolution of the 
sfermion mass parameters including the right-handed neutrino sector is compared 
with the evolution if this sector is cut off.  

\vspace{3mm}
\noindent
{\underline{\it mSUGRA vs. GMSB}:} In gauge mediated supersymmetry breaking GMSB 
\cite{gmsb}
the system is characterized by two scales, defined by the vacuum expectation values
of components of the superfield $S$ inducing the symmetry breakdown
in the secluded sector. 
They are related to the masses of the messengers  
which transport the breaking of SUSY from the secluded sector 
to the eigen-world [$M_M \sim$~PeV within wide margins], and the mass scale 
$\Lambda$ setting the size of the gaugino 
and scalar masses. Modulo threshold factors, the gaugino masses $M_i$ and the 
scalar masses $M_{\tilde j}$,
generated by messenger and gauge field induced loops, can be written in compact
form,
\begin{eqnarray}
M_i(M_M) &\approx&  N_M \, \alpha_i(M_M) \, \Lambda \\
M_{\tilde \jmath}^2(M_M) &\approx& 2 N_M \sum_{i=1}^{3} k_i \, C^j_i\,
        \alpha_i^2(M_M) \, \Lambda^2
\end{eqnarray}
with $N_M$ denoting the multiplicity of messenger multiplets, 
$k_i$ and $C_i^j$ 
being group factors, and the $\alpha_i$'s are 
the three gauge couplings. Ratios of 
scalar masses, $M^2_{\tilde \jmath}(M_M) /$ $M^2_{\tilde \jmath'}(M_M)$ depend only on group factors 
and gauge couplings so that they can be predicted uniquely in GMSB. 

\begin{figure}[tb]
\setlength{\unitlength}{1mm}
\begin{center}
\begin{picture}(70,55)
\put(-1,-7){
\mbox{\includegraphics[height=6cm,width=7cm,angle=0,viewport=10 290 270 520]{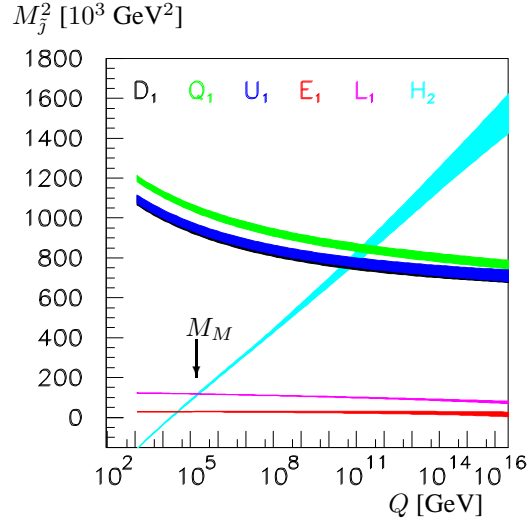}}}
\put(0,55){\mbox{$M^2_{\tilde\jmath}$~[$10^3$ GeV$^2$]}}
\put(50,-10){\mbox{$Q$~[GeV]}}
\put(24.5,13){\vector(0,-1){5}}
\put(23,13.5){$M_M$}
\end{picture}
\end{center}
\caption{{\bf GMSB:} {\it  
Evolution of first-generation slepton mass parameters and Higgs mass parameters
$M_{H_2}^2$. The point probed, SPS\#8 is characterized by the parameters $M_M =
200$ TeV, $\Lambda = 100$ TeV, $\tan\beta = 15$, and
$\rm sign (\mu) = (+)$. [The widths of the bands indicate the 1$\sigma$ CL.]}}
\label{fig:gmsb}
\end{figure} 
The evolution of the scalar mass parameters is shown in Fig.~\ref{fig:gmsb}. The bands 
of the slepton $\tilde {L}$-doublet and the second Higgs doublet $H_2$, which carry the 
same moduli of standard-model charges, cross at the scale $M_M$. The two scales $M_M$ 
and $\Lambda$, and the messenger multiplicity $N_M$ can be extracted from the spectrum 
of the gaugino and scalar particles. For the reference point analyzed in the 
Fig.~\ref{fig:gmsb}, the following accuracies can be obtained:
\begin{eqnarray}
\Lambda &=& (1.01 \pm 0.03) \cdot 10^2 {\rm \ TeV} \nonumber \\
M_M &=& (1.92 \pm 0.24) \cdot 10^2 {\rm \ TeV} \nonumber \\
N_M &=& 0.978 \pm 0.056 \nonumber
\end{eqnarray}
Comparing the two figures representative for the evolution of the scalar mass
parameters, it is manifest that mSUGRA will not be confused with GMSB so long
as the messenger scale does not move out of the PeV range to the Planck scale
-- which would be a  {\it contradictio in origine}.

\section{String Effective Field Theory}
\noindent
Among the most exciting candidates for a comprehensive theory of matter
and interactions rank superstring theories. We will
summarize results obtained for
a string effective field theory in four
dimensions based on orbifold compactification of the 10-dimensional
heterotic superstring~\cite{Binetruy:2001md}.
SUSY breaking is generated non-perturbatively in this approach, mediated by a Goldstino field that is the superposition
of the dilaton field $S$ and the moduli fields $T$ 
[all moduli fields assumed
to be of identical structure]:
\begin{equation}
G = S \sin\theta + T \cos\theta
\end{equation}
Universality is generally broken in such a scenario by a set of non-universal
modular weights $n_j$ that determine the coupling of $T$ to the SUSY matter
fields $\Phi_j$.

The gaugino and scalar mass parameters can be expressed to leading order
by the gravitino mass $m_{3/2}$, the vacuum value $\langle S \rangle$, 
the mixing parameter
$\sin\theta$, and the modular weights $n_j$:
\begin{eqnarray}
M_i &=&  - g_i^2 m_{3/2} {\langle S \rangle} {\sqrt{3} 
           \sin \theta} + \cdots \nonumber \\
M_{\tilde j}^2 &=& m^2_{3/2} \left(
  1 + n_j \cos^2 \theta \right) + \cdots 
\end{eqnarray}
\noindent
while in next-to-leading order, indicated by the ellipses,
the vacuum
value $\langle T \rangle$ and the Green-Schwarz parameter $\delta_{\rm GS}$ 
are included. The
system is completed by relations between the universal gauge coupling
$\alpha(M_{\rm string})$ at the string scale $M_{\rm string}$ 
and the [slightly
non-universal] gauge couplings $\alpha_i(M_{\rm GUT})$ at the 
SU(5) unification scale $M_{\rm GUT}$:
\begin{equation}
\alpha^{-1}_i(\rm M_{\rm GUT}) = \alpha^{-1}({\rm M_{\rm string}}) + 
\Delta \alpha^{-1}_i[n_j]
\end{equation}
The small deviations of the gauge couplings from universality at
the GUT scale are accounted for by string loop effects transporting the
couplings from the universal string scale to the GUT scale.
The gauge coupling at $M_{\rm string}$ is related to the dilaton field, 
$g_s^2 = 1/{\langle S \rangle}$.

\begin{figure}[tb]
\setlength{\unitlength}{1mm}
\begin{center}
\begin{picture}(70,55)
\put(-1,-10){
\mbox{\includegraphics[height=7cm,width=7cm,angle=0]{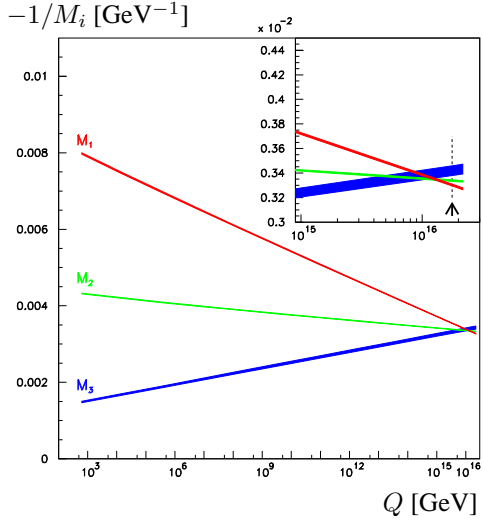}}}
\put(0,55){\mbox{$-1/M_i$~[GeV$^{-1}$]}}
\put(50,-10){\mbox{$Q$~[GeV]}}
\end{picture}
\end{center}
\caption{{\bf String Scenario:} {\it  
Gaugino mass parameters for the heterotic string [the insert 
expands on the breaking of universality at the GUT scale].  The 
point probed is characterized by the
parameters  $m_{3/2}=180$~{\normalfont GeV}, $\delta_{GS}=0$,
 $\langle S \rangle = 2$, $\langle T \rangle =14 \, m_{3/2}$, 
 $\sin^2 \theta=0.9$,
  $\tan \beta = 10$, $\mathrm{sign}(M_2 \mu) = (+)$,
   $n_{L_i} = -3$, $n_{E_i} = -1$, $n_{H_1} =n_{H_2}=-1$,
     $n_{Q_i} = 0$, $n_{D_i} = 1$ and $n_{U_i} = -2$. 
[The widths of the bands indicate the $1 \, \sigma$ CL.]
}}
\label{fig:string}
\end{figure} 

\begin{table}[tb]
\begin{center}
\begin{tabular}{|c||c|r@{$\;$}c@{$\;$}l|}
\hline
Parameter           & Ideal & \multicolumn{3}{c|}{Reconstructed} \\ \hline\hline
$m_{3/2}$           &  180  &     179.9 & $\pm$ & 0.4 \\
$\langle S \rangle$ &   2   &      1.998 & $\pm$ & 0.006 \\
$\langle T \rangle
  /m_{3/2}$ &  14   &      14.6 & $\pm$ & 0.2 \\
$\sin\theta$        &  0.949&      0.948 & $\pm$ & 0.001 \\
$g_s^2$             & 0.5   &      0.501 & $\pm$ & 0.002 \\
$\delta_{GS}$       &   0   &      0.1 & $\pm$ & 0.4 \\ \hline
$n_L$               &  -3   &      -2.94 & $\pm$ & 0.04 \\
$n_E$               &  -1   &     -1.00 & $\pm$ & 0.05 \\
$n_Q$               &   0   &     0.02 & $\pm$ & 0.02 \\
$n_U$               &  -2   &     -2.01 & $\pm$ & 0.02 \\
$n_D$               &  +1   &      0.80 & $\pm$ & 0.04 \\
$n_{H_1}$           &  -1   &      -0.96 & $\pm$ & 0.06 \\
$n_{H_2}$           &  -1   &      -1.00 & $\pm$ & 0.02 \\ \hline
$\tan \beta$        &  10   &      10.00 & $\pm$ & 0.13 \\ 
\hline
\end{tabular}
\end{center}
\caption[]{\it Comparison of the experimentally reconstructed values with the
               ideal fundamental parameters in a specific example 
    for a string  effective field theory; masses in {\normalfont GeV}.} 
\label{tab:parameters_string2}
\end{table}
The evolution of the gaugino masses in such a scenario is illustrated in
Fig.~\ref{fig:string}, with the crucial high-scale region expanded 
in the insert. Relevant parameters constructed from an overall-fit to
couplings and masses are collected in Tab.~\ref{tab:parameters_string2}. 
It turns out that the ideal values, from which the experimental
input observables were derived, can indeed be extracted from the data
collected at high-energy hadron- and lepton-colliders that will allow to perform
the high-precision measurements required for this purpose.

\section{Conclusions}
\noindent
In this summary report we have demonstrated that, based on future
high-precision data from e$^+$e$^-$ linear colliders, TESLA in particular, and
combined with results from LHC, and later CLIC, the fundamental supersymmetry
parameters can be reconstructed at the high scale, GUT or Planck, in practice.
The bottom-up approach of evolving the parameters from the low-energy scale to
the high scale by means of renormalization group techniques provides us with a
transparent picture in a region where gravity is linked to particle physics,
and superstring theory becomes effective directly. We have exemplified this --
truly exciting -- observation in two ways explicitly, for minimal supergravity
theories, and for a string effective field theory based on orbifold
compactification of the heterotic string. We could demonstrate that the
effective string parameters can indeed be reconstructed from high-precision
high-energy experiments at hadron- and lepton-colliders.

\vspace{1em}
\noindent {\bf Acknowledgements:} This work was supported in part by the
  European Commision 5th framework under contract HPRN-CT-2000-00149 and
  the Polish-German LC project No. POL 00/015. W.~P.~is
  supported by the ''Erwin Schr\"odinger fellowship No.~J2095" of
the ''Fonds zur F\"orderung der wissenschaftlichen Forschung" of Austria
FWF and partly by the Swiss ''Nationalfonds".

\end{document}